\documentclass[preprint]{imsart}
\usepackage{amsmath,amssymb,amsfonts,graphicx,amsthm,nicefrac,mathtools,bm}
\usepackage[numbers]{natbib}
\usepackage[english]{babel}
\usepackage{times}
\usepackage[T1]{fontenc}
\usepackage{bbm}
\usepackage{color}
\usepackage{xspace}
\usepackage{rotating,multirow}

\startlocaldefs

\theoremstyle{definition}

\theoremstyle{remark}

\makeatletter
\newtheorem*{rep@theorem}{\rep@title}
\newcommand{\newreptheorem}[2]{%
\newenvironment{rep#1}[1]{%
 \def\rep@title{#2 \ref{##1}}%
 \begin{rep@theorem}}%
 {\end{rep@theorem}}}
\makeatother
\newreptheorem{theorem}{Theorem}
\newreptheorem{lemma}{Lemma}
\newreptheorem{corollary}{Corollary}
\newreptheorem{proposition}{Proposition}

\newcommand{\ignore}[1]{}

\newcommand{\nothere}[1]{}

\newcommand{\bmX}{\mbox{\boldmath $X$}}
\newcommand{\bmR}{\mbox{\boldmath $R$}}

\newcommand{\bmH}{\mbox{\boldmath $H$}}
\newcommand{\bmS}{\mbox{\boldmath $S$}}

\newcommand{\bmx}{\mbox{\boldmath $x$}}

\newcommand{\bmI}{\mbox{\boldmath $I$}}

\newcommand{\bme}{\mbox{\boldmath $e$}}

\newcommand{\bmbeta}{\mbox{\boldmath $\beta$}}
\newcommand{\bmOmega}{\mbox{\boldmath $\Omega$}}
\newcommand{\bmPsi}{\mbox{\boldmath $\Psi$}}

\newcommand{\bmalpha}{\mbox{\boldmath $\alpha$}}

\newcommand{\bmgamma}{\mbox{\boldmath $\gamma$}}

\newcommand{\ba}{\begin{array}}
\newcommand{\ea}{\end{array}}

\newcommand{\bea}{\begin{eqnarray*}}
\newcommand{\eea}{\end{eqnarray*}}

\newcommand{\beaa}{\begin{eqnarray}}
\newcommand{\eeaa}{\end{eqnarray}}

\endlocaldefs

\begin{document}

\begin{frontmatter}

\title{Instrumental Variable with Competing Risk Model}
\runtitle{Instrumental Variable with Competing Risk Model}


\author{\fnms{Cheng} \snm{Zheng}\ead[label=e1]{zhengc@uwm.edu}}
\address{Joseph. J. Zilber School of Public Health, University of Wisconsin-Milwaukee, Milwaukee, WI\\ \printead{e1}}

\and
\author{\fnms{Ran} \snm{Dai}}
\address{Department of Statistics, University of Chicago, Chicago, IL}

\and
\author{\fnms{Parameswaran} \snm{Hari}}
\address{Division of Hematology and Oncology, Medical College of Wisconsin, Milwaukee, WI}

\and
\author{\fnms{Mei-Jie} \snm{Zhang}}
\address{Division of Biostatistics, Medical College of Wisconsin, Milwaukee, WI}

\runauthor{C. Zheng \emph{et al.}}

\begin{abstract}
In this paper, we discuss causal inference on the efficacy of a treatment or medication on a time-to-event outcome with competing risks. Although the treatment group can be randomized, there can be confoundings between the compliance and the outcome. Unmeasured confoundings may exist even after adjustment for measured covariates. Instrumental variable (IV) methods are commonly used to yield consistent estimations of causal parameters in the presence of unmeasured confoundings. Based on a semi-parametric additive hazard model for the subdistribution hazard, we propose an instrumental variable estimator to yield consistent estimation of efficacy in the presence of unmeasured confoundings for competing risk settings. We derived the asymptotic properties for the proposed estimator. The estimator is shown to be well performed under finite sample size according to simulation results. We applied our method to a real transplant data example and showed that the unmeasured confoundings lead to significant bias in the estimation of the effect (about 50\% attenuated).
\end{abstract}



\end{frontmatter}

\section{Introduction}
In clinical trials, randomization of the treatment group can only guarantee an unbiased estimation of the intent to treat (ITT) effect. In most cases, the true efficacy of a new drug or treatment is of more interest. However, to yield a causal interpretation of the association between the drug intake/treatment and the effect, we need to remove potential confoundings between the compliance and the outcome. The assignment of treatments and the compliance to the treatments are highly dependent on the interactions between the physician and the patients. It is often hard to make all confoundings controlled by adjusting measured covariates. One important technique to handle the confounding issues in causal inference is the use of instrumental variables (IV) \cite{r4}, which was first introduced in econometric literatures and then applied in the field of biomedical sciences \cite{r11, r2}. The concept of instrumental variable had a srong influence on causal inference studies and inspired methods such as mediation analysis \cite{r28} and Mendelian randomization \cite{r7}. By definition, an instrumental variable is a variable that is related to the exposure variable of interest but not directly linked to the outcome variable. For example, researchers used children's birthdates as  an instrument to assess the effect of education on earning \cite{r1} since this variable is correlated with years of education (the exposure variable) but not directly associated with final earnings (the response).

In this paper, we are interested in estimating the efficacy of rituximab drug on diffuse large B Cell Lymphoma (DLBCL) patients. The decision of whether to use the drug depends on complicated factors and we are not able to fully control the treatment process and remove the confoundings correspondingly. Since the FDA approval of rituximab affects its use frequencies in clinical practice and the calendar time is highly associated with the use of rituximab \cite{r8}, we can consider the calendar time as an instrumental variable. This variable has an influence on the patients' usage of rituximab, but does not change rituximab's drug effect.

In linear models, two commonly used instrumental variable estimators are two-stage least squares (TSLS) \cite{r25} and generalized method of moments (GMM) \cite{r10}. The first method is easier to implement using the existing regression softwares while the second is potentially more efficienct when there are many instruments to use and when the model is over-identified. To use instrumental variables in the context of time-to-event data with censoring, Robins and Tsiatis \cite{r17} proposed an IV method to handle noncompliance with the requirement that the censoring time is known (administrative censoring), which was extended to proportional hazard model by Loeys and Goetghebeur\cite{r15}. IV methods that rely on other parametric model assumptions were also proposed \cite{r5}. Recently, Li \emph{et al.} \cite{r13} proposed an IV based two-stage estimator for a semi-parametric additive hazard model\cite{r14} and Chan \cite{r26} extended the TSLS method to allow exposure dependent censoring.

Competing risks data are commonly encountered in biomedical research. In competing risks data analysis, we are often interested in studying the covariate effects on the cumulative incidence function (CIF) of one specific cause of failure. Regression modeling, such as Cox proportional hazards model, was considered as a standard approach to model the cause-specific hazard function for each cause of failure \cite{r6,r23,r18,r19}. It is hard to summarize the covariate effects on CIF of a particular type of failure since modeling of the cause specific hazards would result in a complex nonlinear relationship for the CIFs. To bypass this difficulty, methods that directly model the subdistribution hazard function of a specific cause were proposed\cite{r9,r24,r29}. However there is a gap in handling confounding issues in competing risk data analysis. Scheike \emph{et al.} \cite{r20,r22} proposed a class of flexible regression models, including multiplicative model and additive model for competing risks data through a direct binomial regression modeling approach. The idea for the construction of this model is natural and it is easy to accommodate the model assumptions with additive model in IV analysis for the competing risks data. To fill the gap, we extend the method of Li \emph{et al.} \cite{r13} to the competing risks data analysis using an additive hazard subdistribution model.


In section 2, we present our proposed model followed by detailed assumptions and estimation procedures. In section 3, the theoretical asymptotic results are provided. In section 4, we show extensive simulation results and discuss the performance of our proposed estimator for finite sample size compared with the simple additive hazard model fitting without using IV. In section 5, we apply our proposed method to a real data example and compare our results with the original fitting without using IV. In section 6, we provide more discussions of the model and its extensions. The detailed proofs can be found in the appendix.

\section{Model}
\subsection{Notation}
Consider a competing risk study with $K$ types of failures. Let $T$ and $C$ denote the failure time and censoring time. $\epsilon\in \{1,\cdots, K\}$ indicates the cause of failure. Consider right censoring, we observe a composite outcome, $T^{\ast}=\min(T,C)$ and $\delta=I(T\leq C)$. Denote $X_e$ as exposure of primary interest. Denote $\bmX_o\in \bmR^{p}$ and $\bmX_u\in \bmR^{q}$ as the observed and unobserved exogenous risk factors. Denote $X_I$ as the instrumental variable. Here we begin with one dimension of $X_e$ and one dimension of $X_I$ which fits our motivation example. The general form with more predictors of interest and over-identified model are discussed in later sections. Without loss of generality, we assume that all covariates are centered at zero. Based on our notation above, our observed data sets are $\{T^{\ast},\delta,\delta\epsilon, X_e, \bmX_o, X_I\}$. Let $\bmX=\{X_e, \bmX_o, \bmX_u, X_I\}$, we model the cause specific cumulative incidence function as $F_1(t|\bmX)=P(T\leq t, \epsilon=1|\bmX)$ for the subgroup with covariate $\bmX$. For notation simplicity, we treat $\bmX$ as time-independent. Our method can be extended to the external time-varying covariate process (i.e. interactions between time and baseline covariates) when the covariate process is available after the event occurs. We denote $N_i(t)=I(T_i\leq t, \epsilon_i=1)$, $Y_i(t)=1-N_i(t^-)$. Since $N_i(t)$ and $Y_i(t)$ are not observable for all time points of censored subjects, we introduce $r_i(t)=I(C_i\geq T_i\wedge t)$, then both $r_i(t)N_i(t)$ and $r_i(t)Y_i(t)$ are computable for all time points.

\subsection{Structural Additive Model for Sub-distribution Hazard}
Given the important feature that for additive hazard model, the effect is collapsible over unmeasured effects, we consider an IV analysis based on an additive subdistribution hazard model. Suppose we are interested in cause 1, then we model the subdistribution hazard for cause 1 as
$$
\lambda_1(t|\bmX)=-\frac{d}{dt}\log\left\{1-F_1(t|\bmX)\right\},
$$
our model will be in the form of
$$
\lambda_1(t|\bmX)=h_0(t)+\beta_eX_e+\bmbeta_o^\textsf{T} \bmX_o + \bmbeta_u^\textsf{T} \bmX_u,
$$
where $\beta_e \in \bmR$, $\bmbeta_o\in \bmR^p$ and $\bmbeta_u\in \bmR^q$.

Thus, we have
\begin{align*}
F_1(t|\bmX)&=1-\exp\left\{-\int_0^{t}\lambda_1(s|\bmX)ds\right\}\\
&=1-\exp\left\{-H_0(t)-\beta_eX_e t-\bmbeta_o^\textsf{T} \bmX_o t-\bmbeta_u^\textsf{T} \bmX_u t\right\},
\end{align*}
where $\displaystyle H_0(t)=\int_0^{t}h_0(s)ds$.
So
\begin{align*}
\lambda_1(t|X_e,\bmX_o)&=-\frac{d}{dt}\log\left \{1-F_1(t|X_e,\bmX_o)\right\}\\
&=-\frac{d}{dt}\log\left[1-E\{F_1(t|\bmX)|X_e,\bmX_o\}\right]\\
&=-\frac{d}{dt}\log\left\{\int\exp\left(-H_0(t)-\beta_e X_e t-\bmbeta_o^\textsf{T} \bmX_o t-\bmbeta_u^\textsf{T} \bmX_u t\right)dF(\bmX_u|X_e,\bmX_o)\right\}\\
&=-\frac{d}{dt}\log\left\{\exp\left(-H_0(t)-\beta_e X_e t-\bmbeta_o^\textsf{T} \bmX_o t\right)\int\exp\left(-\bmbeta_u^\textsf{T} \bmX_ut\right ) dF(\bmX_u|X_e,\bmX_o)\right\}\\
&=h_0(t)+\beta_e X_e +\bmbeta_o^\textsf{T} \bmX_o -\frac{d}{dt}\log\left [E\left\{\exp\left(-\bmbeta_u^\textsf{T} \bmX_u t\right)|X_e,\bmX_o\right\}\right].
\end{align*}
We assume $E(\bmX_u|\bmX_o, X_e)=\bmalpha_e X_e+\bmalpha_o^\textsf{T} \bmX_o$ and $\mbox{Var}(\bmX_u|\bmX_o, X_e)=\sigma_u^2 \bmI$, where
$\bmalpha_e$ is a $(q\times 1)$-vector and $\bmalpha_o^\textsf{T}$ is a $(q\times p)$-matrix.
Then by second order approximation, the above expression can be further simplified by the following approximation,
\begin{align*}
\tilde{h}_0(t)+\beta_e X_e+ \bmbeta_o^\textsf{T} X_o +\bmbeta_u^\textsf{T} \left(\bmalpha_e X_e + \bmalpha_o^\textsf{T} \bmX_o\right)
 =\tilde{h}_0(t)+\left (\beta_e +\bmbeta_u^\textsf{T} \bmalpha_e\right )X_e+\left (\bmbeta_o+\bmbeta_u^\textsf{T} \bmalpha_o^\textsf{T}\right )\bmX_o
\end{align*}
where $\tilde{h}_0(t)=h_0(t)+\bmbeta_u^2\sigma_u^2t$ and $\bmbeta_u^2=\bmbeta_u^\textsf{T} \bmbeta_u$. Since $\bmX_u$ is confounding, $\bmalpha_e\neq \mbox{\boldmath $0$}$, thus the regression estimator for the treatment effect based on observed variables will be biased. The above approximation holds exactly when $\bmX_u=\bmalpha_e X_e + \bmalpha_o^\textsf{T} \bmX_o + \bme_u$, with $\bme_u$ normally distributed and is independent of $X_e$ and $\bmX_o$.

In order to obtain a consistent estimator of $\beta_e$, we need to use the instrumental variable $X_I$. Assuming that $X_e$ can be modeled as:
\begin{align*}
X_e=\gamma_0+\gamma_IX_I+\bmgamma_o^\textsf{T} \bmX_o + \bmgamma_u^\textsf{T} \bmX_u + e.
\end{align*}

We present detailed assumptions required for the instrumental variable analysis in the following section.
We denote $R_i(t)=r_i(t)G(t)/G(T_i\wedge t)$, $Y^{\ast}_i(t)=R_i(t)Y_i(t)$ and $N^{\ast}_i(t)=R_i(t) N_i(t)$. Denote $\hat{R}_i(t)=r_i(t)\hat{G}(t)/\hat{G}(T_i\wedge t)$, where $G(t)=P(C\geq t)$ and $\hat{G}(t)$ is estimated from either Kaplan-Meier estimator when assuming random censoring or from a specific regression model when ${G}_i(t)=P(C\geq t|\bmX_{oi},X_{Ii})$ depends on $\bmX_o$ and $X_I$. For simplicity, only Kaplan-Meier estimate for the censoring distribution is considered through out the paper. In practice we may consider using a stratified Kaplan-Meier estimate when the censoring distribution is dependent of some observed covariates. Denote $\hat{Y}_i(t)=\hat{R}_i(t)Y_i(t)$ and $\hat{N}_i(t)=\hat{R}_i(t)N_i(t)$. We assume that censoring distribution $C$ is conditionally independent of $(T,\epsilon)$ conditioning on variables $X_I$ and $\bmX_o$ (Assumption 6), which is stronger than traditional assumption that $C$ is conditionally independent of $(T,\epsilon)$ conditioning on variables $X_I$, $X_e$ and $\bmX_o$. This assumption is needed since $X_e$ will not be included in the model for the second stage of the two-stage estimator. Other assumptions are parallel to those required for traditional additive hazard model. Thus, when Assumption 6 holds, the two models' results are comparable.

Now we propose the two-stage estimation procedure. In the first stage, we fit a model of $X_e$ on $X_I$ and $\bmX_o$ to obtain $\hat{\gamma}_0$, $\hat{\gamma}_I$ and $\hat{\bmgamma}_o^\textsf{T}$. Then compute $\hat{X}_{ei}=\hat{\gamma}_0+\hat{\gamma}_I X_{Ii}+\hat{\bmgamma}_o^\textsf{T} \bmX_{oi}$. In the competing risk model, we regress on $\hat{X}_{ei}$ and $\bmX_{oi}$ rather than $X_{ei}$ and $\bmX_{oi}$ to obtain consistent estimator $\hat{\beta}_e$. For the first stage estimator, by the result of ordinary least square, we can estimate $\hat{\bmgamma}=(\hat{\gamma}_0,\hat{\gamma}_I,\hat{\bmgamma}_o)^\textsf{T}$ with
\begin{eqnarray*}
\hat{\bmgamma}=[\bmX_{Io}^\textsf{T} \bmX_{Io}]^{-1}[\bmX_{Io}^\textsf{T} \bmX_e],
\end{eqnarray*}
where $\bmX_{Io}=[\mbox{\boldmath $1$}, \bmX_I, \bmX_o]$ is a $n\times(p+2)$ matrix. It is obvious that $\hat{\gamma}_0$, $\hat{\gamma}_I$, $\hat{\bmgamma}^\textsf{T}_o$ are consistent estimators of $\gamma_0$, $\gamma_I$, $\bmgamma_o^\textsf{T}$. Thus define $\tilde{\bmX}_e=\bmX_{Io}\bmgamma$, then we have $\hat{\bmX}_e=\bmX_{Io}\hat{\bmgamma}=\bmX_{Io}[\bmX_{Io}^\textsf{T} \bmX_{Io}]^{-1} \bmX_{Io}^\textsf{T} \bmX_e=\bmH_{Io} \bmX_e$, where $\bmH_{Io}=\bmX_{Io}[\bmX_{Io}^\textsf{T} \bmX_{Io}]^{-1} \bmX_{Io}^\textsf{T}$ is the projection matrix on $\bmX_{Io}$. The ordinary linear regression result guarantees that $\hat{\bmX}_e$ converge to $\tilde{\bmX}_e$ uniformly in probability in a compact set of $\bmX_{Io}$.

In the second stage, we fit a subdistribution additive hazard model on covariates $\hat{X}_{ei}$, $\bmX_{oi}$. To fit this model, we applied the estimator proposed by Sun \emph{et al.} \cite{r24} and used the R programming method developed by Scheike and Zhang \cite{r21}. We use estimated $\hat{X}_{ei}$ as our final estimator for $\beta_e$. To see why this works, we make the following brief derivation:
\begin{align*}
\lambda_1(t|\tilde{X}_{ei},\bmX_{oi})&=-\frac{d}{dt}\log\left \{1-F_1\left(t|\tilde{X}_{ei},\bmX_{oi}\right)\right\}\\
 &=-\frac{d}{dt}\log\left\{1-\int F_1 \left (t|\tilde{X}_{ei},\bmX_{oi},\bmX_{ui},e_i\right) dF\left(\bmX_{ui},e_i|\tilde{X}_{ei},\bmX_{oi}\right)\right\}.
\end{align*}
According to the regularity conditions (see assumption 1, 4 and 5) and the fact that $F\left(\bmX_{ui},e_i|\tilde{X}_{ei},\bmX_{oi}\right)=F\left(\bmX_{ui},e_i|X_{Ii},\bmX_{oi}\right)$ it can be further written as
\begin{align*}
&-\frac{d}{dt} \log\left\{1-\int F_1\left(t|\tilde{X}_{ei},\bmX_{oi},\bmX_{ui},e_i\right)dF(\bmX_{ui},e_i|\tilde{X}_{ei},\bmX_{oi})\right\}\\
&=-\frac{d}{dt} \log\left[\int\exp\left\{- H_0(t) -\beta_e\left (\tilde{X}_{ei} + \bmgamma_{u}^\textsf{T} \bmX_{ui} +e_i\right ) t -\bmbeta_o^\textsf{T} \bmX_{oi} t -\bmbeta_u^\textsf{T} \bmX_{ui} t \right\} dF(\bmX_{ui},e_i|X_{Ii},\bmX_{oi})\right]\\
&=-\frac{d}{dt} \log\left[\int \exp\left \{- H_0(t) -\beta_e\tilde{X}_{ei} t -\beta_e e_i t -\bmbeta_o^\textsf{T} \bmX_{oi} t -(\bmbeta_u^\textsf{T} +\beta_{ei} \bmgamma_u^\textsf{T} )\bmX_{ui} t\right \}dF\left(\bmX_{ui},e_i|X_{Ii},\bmX_{oi}\right)\right]\\
&=-\frac{d}{dt} \log\left[ \exp\left\{- H_0(t) -\beta_e \tilde{X}_{ei} t-\bmbeta_o^\textsf{T} \bmX_{oi} t\right\}\int \exp\left\{- \beta_{ei} e_i t - \left (\bmbeta_u^\textsf{T} + \beta_e \bmgamma_u^\textsf{T}\right ) \bmX_{ui} t\right\}dF(X_{ui},e_i|X_{Ii},\bmX_{oi})\right]\\
&=h^{\ast}_0(t)+\beta_e \tilde{X}_{ei} + \bmbeta_o^\textsf{T} \bmX_{oi},
\end{align*}
where $\displaystyle h^{\ast}_0(t)=h_0(t) - \frac{d}{dt}\log\left[\int \exp\left\{-\beta_e e_i t - \left(\bmbeta_u + \beta_e \bmgamma_u\right )^\textsf{T} \bmX_{ui} t\right\} dF(X_{ui},e_i|X_{Ii},\bmX_{oi})\right]$. Under assumption 4, with second order approximation, we have
\begin{eqnarray*}
&&\frac{d}{d\tilde{X}_e}\left\{-\frac{d}{dt}\log\left[\int \exp\left\{-\beta_e e_i t - \left(\bmbeta_u + \beta_e \bmgamma_u\right )^\textsf{T} \bmX_{ui} t\right\} dF(X_{ui},e_i|X_{Ii},\bmX_{oi})\right]\right\}\\
&\approx &\frac{d}{d\tilde{X}_e}\left[E[\bmgamma_u^\textsf{T}\bmX_u+e|X_{I},\bmX_{o}]\beta_e+Var[\bmgamma_u^\textsf{T}\bmX_u+e|X_{I},\bmX_{o}]\beta_e^2t\right].
\end{eqnarray*} 
The approximation hold exactly when $\bmgamma_u^\textsf{T}\bmX_u+e$ follow normal distribution and the quantity on the right hand side equal to 0 when $\bmgamma_u^\textsf{T}X_u+e$ is mean 0 with constant variance, which is gauranteed by assumption 1.

Denote $\bmbeta^\textsf{T}=(\beta_e,\bmbeta_o^\textsf{T})$. We have the following unbiased estimating equations
\begin{align*}
0&=\sum_{i=1}^n R_i(t)Y_i(t) \left\{dR_i(t) N_i(t) - R_i(t)Y_i(t) \bmbeta^\textsf{T} \left(\tilde{X}_{ei},\bmX_{oi}\right ) dt -R_i(t)Y_i(t) dH_0(t)\right\} \\
0&=\sum_{i=1}^n \int_{0}^{\tau} \left(\tilde{X}_{ei},\bmX_{oi}\right )R_i(t)Y_i(t) \left\{dR_i(t) N_i(t) - R_i(t)Y_i(t) \bmbeta^\textsf{T} \left(\tilde{X}_{ei},\bmX_{oi}\right ) dt -R_i(t)Y_i(t) dH_0(t)\right\}.
\end{align*}
Replace the unknown quantity $R_i(t)$ and $\tilde{X}_{ei}$ by their consistently estimated version $\hat{R}_i(t)$ and $\hat{X}_{ei}$, we can solve the following estimation equation to obtain $\hat{\bmbeta}^\textsf{T}=(\hat{\beta}_e,\hat{\bmbeta}^\textsf{T}_o)$.
Denote $\bmX_{IOE}=\left[\ba{c}\bmX_e^\textsf{T} \bmX_{Io}\left (\bmX_{Io}^\textsf{T}\bmX_{Io}\right)^{-1}\\(0_p,0_p,I_p) \ea\right]$, which is a $(p+1)\times (p+2)$ matrix.
With known weighting function $w_i(t)$,we have
\begin{align*}
0&=\sum_{i=1}^n w_i(t)\hat{Y}_i(t)\left\{ d\hat{N}_i(t)-\hat{Y}_i(t)\bmbeta^\textsf{T} \bmX_{IOE}X_{Ioi} dt -\hat{Y}_i(t)dH_0(t)\right\} \\
0&=\sum_{i=1}^n\int_{0}^{\tau}w_i(t)\hat{Y}_i(t)\bmX_{IOE}X_{Ioi}\left\{ d\hat{N}_i(t)-\hat{Y}_i(t) \bmbeta^\textsf{T} \bmX_{IOE}\bmX_{Ioi} dt-\hat{Y}_i(t)dH_0(t)\right\}.
\end{align*}
From the first equation, we obtain
\[
d\hat{H}_0(t)=\frac{\sum_{i=1}^nw_i(s)\hat{Y}_i(t)\left\{d\hat{N}_i(t)-\hat{Y}_i(t)\bmbeta^\textsf{T} \bmX_{IOE}\bmX_{Ioi}\right\}}{\sum_{i=1}^n w_i(t)\hat{Y}^2_i(t)}.
\]
Plug the first equation solution into the second, we have
\begin{align*}
0=&\sum_{i=1}^n\int_{0}^{\tau} \left[\bmX_{IOE}\left\{\bmX_{Ioi} -\frac{\sum_{j=1}^n \bmX_{Ioj} w_j(t)\hat{Y}^2_j(t)}{\sum_{j=1}^n w_j(t)\hat{Y}^2_j(t)}\right\} \right] \\ &\times w_i(t)\hat{Y}_i(t)\left\{d\hat{N}_i(t)-\hat{Y}_i(t)\bmbeta^\textsf{T} \bmX_{IOE}\bmX_{Ioi} dt\right\}.
\end{align*}
So we have
\begin{align*}
\hat{\bmbeta}&=\left(n^{-1} \sum_{i=1}^n \int_{0}^{\tau}\left[\bmX_{IOE}\left\{\bmX_{Ioi} - \frac{\sum_{j=1}^n \bmX_{Ioj}w_j(t) \hat{Y}^2_j(t)}{\sum_{j=1}^n w_j(t)\hat{Y}^2_j(t)}\right\}\right]^{\otimes 2}w_i(t)\hat{Y}^2_i(t)dt \right)^{-1}\\
&\times \left(n^{-1}\sum_{i=1}^n \int_{0}^{\tau}\left[\bmX_{IOE}\left\{\bmX_{Ioi} - \frac{\sum_{j=1}^n \bmX_{Ioj}w_j(t)\hat{Y}^2_j(t)}{\sum_{j=1}^n w_j(t)\hat{Y}^2_j(t)}\right\}\right] w_i(t)\hat{Y}_i(t)d\hat{N}_i(t)\right)\\
&=\bmS_{2n}^{-1}(\hat{\bmgamma},\hat{\Lambda}_c) \bmS_{1n}(\hat{\bmgamma},\hat{\Lambda}_c)
\end{align*}
where $a^{\otimes 2}=aa^\textsf{T}$.

The baseline hazard is not identifiable due to the confounding of $\bmX_u$, i.e., we can estimate $H^{\ast}_0(t)=\int_0^t h^{\ast}_0(s)ds$ but we are not able to estimate $H_0(t)$. However, we are still able to estimate covariate specific subdistribution cumulative incident function since it can be represented by $h^{\ast}_0(t)$ instead of $h_0(t)$, then we have $F_1(t|X_e,\bmX_o)=1- \exp\left (-H_0^{\ast}(t) -t\beta_e X_e - t\bmbeta_o^\textsf{T} \bmX_o\right)$. We can obtain $\hat{F}_1(t|X_e,\bmX_o)=1- \exp\left(-\hat{H}_0^{\ast}(t)-t\hat{\beta}_eX_e - t\hat{\bmbeta}_o^\textsf{T} \bmX_o\right )$ with
\begin{eqnarray*}
\hat{H}_0^{\ast}(t)=\int_0^{t}\frac{\sum_{i=1}^nw_i(s)\hat{Y}_i(s)\left\{d\hat{N}_i(s)-\hat{Y}_i(s)\hat{\bmbeta}^\textsf{T} \bmX_{IOE} \bmX_{Ioi}ds\right \}}{\sum_{i=1}^n w_i(s)\hat{Y}^2_i(s)}
\end{eqnarray*}

To handle the technical difficulty of additive model that the estimated hazard might not be positive, we use the approach as suggested for the original additive model \cite{r14} by defining a modified version of estimation $\hat{H}_{mod}^{\ast}(t)=\max_{0\leq s\leq t}\hat{H}_0^*(s)$. Under regularity condition, this modified version of $\hat{H}_{mod}^{\ast}(t)$ is asymptotically equivalent to $\hat{H}_0^{\ast}(t)$ and thus we proof the asymptotics results for $\hat{H}_0^{\ast}(t)$.

\section{Theoretical Results}\label{sec3}

We assume the following regularity conditions hold throughout the paper.

\begin{description}
\item[{\bf Assumption 1:}]\label{ass1} $X_I$, $\bmX_u$ and $\bmX_o$ are mean zero random variables. $\bmgamma_u^\textsf{T} \bmX_u + e$ is uncorrelated with $\bmX_o$ and $X_I$, i.e., $E\left(\bmgamma_u^\textsf{T} \bmX_u + e|\bmX_o, X_I\right)=0$. The variance is constant, i.e.,\\
    $\mbox{Var}\left(\bmgamma_u^\textsf{T} \bmX_u +e|\bmX_o, X_I\right)=\sigma^2$.
\item[{\bf Assumption 2:}]\label{ass2} The instrumental variable $X_I$ is associated with $X_e$ conditional on $\bmX_o$ and $\bmX_u$, i.e., $\alpha_I\neq 0$.
\item[{\bf Assumption 3:}]\label{ass3} $X_I$ is not directly associated with the outcome event, i.e., $X_I$ is independent of $(T, \epsilon, \bmX_u)$ conditional on $\bmX_o$.
\item[{\bf Assumption 4:}]\label{ass4} The error term $e$ has mean 0. $e$ is independent of $X_I$, $\bmX_o$ and $\bmX_u$.
\item[{\bf Assumption 5:}]\label{ass5} $\bmX_o$ and $\bmX_u$ are independent.\\
The assumptions above are parallel to the traditional IV assumptions, with the uncorrelated assumption replaced by independence assumption since our model is nonlinear. Since we are considering the additive hazard subdistribution model, we also need the following assumptions related to the additive hazard model.
\item[{\bf Assumption 6:}]\label{ass6} Independent censoring, i.e. $C$ is conditionally independent of $(T,\epsilon)$, conditional on variables $X_I$ and $\bmX_o$.
\item[{\bf Assumption 7:}]\label{ass7} Covariates $\bmX_o$, $X_I$, $\bmX_u$, $X_e$ have bounded supports and $\bmbeta$ is an interior point of the possible parameter space.
\item[{\bf Assumption 8:}]\label{ass8} When the time period of interest is $[0,\tau]$, we require $P(C\geq t|\bmX_o,X_I)$ to be bounded away from 0 for $t\in [0,\tau]$.
\item[{\bf Assumption 9:}]\label{ass9} $\displaystyle n^{-1}\sum_{i=1}^n \int_{0}^{\tau}\left[\bmX_{IOE}\left\{\bmX_{Ioi}- \frac{\sum_{j=1}^n \bmX_{Ioj}w_j(t)Y^{\ast 2}_j(t)}{\sum_{j=1}^n w_j(t)Y^{\ast 2}_j(t)}\right\}\right]^{\otimes 2} w_i(t)Y^{\ast 2}_i(t)dt$ converge uniformly with probability 1 to a positive definite matrix process $\bmOmega_{\tau}$.
\item[{\bf Assumption 10:}]\label{ass10} $\displaystyle n^{-1}\sum_{i=1}^n \int_{0}^{\tau}\left[\bmX_{IOE}\left\{\bmX_{Ioi}- \frac{\sum_{j=1}^n \bmX_{Ioj}w_j(t)Y^{\ast 2}_j(t)}{\sum_{j=1}^n w_j(t)Y^{\ast 2}_j(t)}\right\}\right]^{\otimes 2} w_i(t)Y^{\ast 2}_i(t)dN_i^*(t)$ converges uniformly with probability 1 to a positive definite matrix process $\bmPsi_{\tau}$.
\item[{\bf Assumption 11:}]\label{ass11}
\bea
&&\frac{\sigma^2\beta_{0e}^2}{n} \sum_{i=1}^n \int_{0}^{\tau} \bmX_{IOE}\left[\{ \bmX_{Ioi} -
\left\{\frac{\sum_{j=1}^n \bmX_{Ioj} w_j(t) Y^{\ast 2}_j(t)}{\sum_{j=1}^n w_j(t)Y^{\ast 2}_j(t)}\right\} \right] \bmX_{Ioi}^\textsf{T}\left(\bmX_{Io}^\textsf{T} \bmX_{Io}\right)^{-1}\bmX_{Ioi}\\
&&\times \left[\bmX_{IOE}\left\{\bmX_{Ioi}-\frac{\sum_{j=1}^n \bmX_{Ioj} w_j(t) Y^{\ast 2}_j(t)}{\sum_{j=1}^n w_j(t)Y^{\ast 2}_j(t)}\right\}\right]^\textsf{T}w_i(t)Y_i^{\ast 2}(t)dt
\eea
converges uniformly with probability 1 to a positive definite matrix process $\Sigma_{\tau}$.
\item[{\bf Assumption 12:}]\label{ass12} For $\tau<\infty$, there exist bounded continuous functions $g(t)$ and $\Gamma(t)$ such that
$$\left\|\int_0^{t}\frac{n^{-1}\sum_{i=1}^nw_i^2(s)Y^{2 \ast}_i(s)dN^{\ast}_i(s)}{[n^{-1}\sum_{i=1}^nw_i(s)Y^{\ast 2}_i(s)]^2}-g(t)\right\|_{\infty}$$ and
$$\left\|\int_{0}^t \bmX_{IOE}\frac{\sum_{j=1}^n \bmX_{Ioj}w_j(s)Y^{\ast}_j(s)dN^{\ast}_j(s)}{\sum_{j=1}^n w_j(s)Y^{\ast 2}_j(s)} -\Gamma(t)\right\|_{\infty}$$ converge to 0 in probability uniformly over $t\in (0,\tau]$.
\end{description}

We provide theoretical results in this section with brief proof outline. The technical details of the proofs can be found in appendix.
\paragraph*{}\textbf{Theorem 1:} Under assumptions 1-11, $\hat{\bmbeta}$ converges to $\bmbeta_0$ with probability 1.\\
\textbf{Proof:} Since $\hat{\bmgamma}$ and $\hat{\Lambda}_c$ converge to $\bmgamma$ and $\Lambda_c$ uniformly and the regularity conditions ensure that the function space is in Glivenko Cantelli class, we have $\hat{\bmbeta}=S_{2n}^{-1}(\hat{\bmgamma},\hat{\Lambda}_c)S_{1n}(\hat{\bmgamma},\hat{\Lambda}_c)$ converges to $\tilde{\bmbeta}=S_{2n}^{-1}(\bmgamma,\Lambda_c)S_{1n}(\bmgamma,\Lambda_c)$ uniformly with probability 1. By the convergence result for the additive subdistribution hazard model, we know that $\tilde{\bmbeta}$ converges to $\bmbeta_0$ with probability 1. Thus $\hat{\bmbeta}$ converges to $\bmbeta_0$ with probability 1.

For simplicity of the variance form, we will choose Kaplan-Meier estimator for $\hat{G}(t)$ and assume random censoring. In practice, if we believe censoring depends on $X_o$ and $X_I$, we can model it via certain semiparametric models. Under the correct specification of such models, the consistency and asymptotic normality results still hold though the variance will have a more complicated form and the Bootstrap will be recommended for.

\paragraph*{}\textbf{Theorem 2:} Under assumptions 1-11, we have
\bea
\sqrt{n}(\hat{\bmbeta}-\bmbeta_0)\rightarrow_d N(0, \Omega_{\tau}^{-1}(\Psi_{\tau}+\Sigma_{\tau})\Omega_{\tau}^{-1})
\eea
with influence function $\Phi_i$ as defined in appendix. The variance can be estimated by using an empirical version of $\hat{\Phi}_i$ of $\Phi_i$ and compute $n^{-1}\sum_{i=1}^n\Phi_i^{\otimes 2}$ to estimate $E\Phi_i^{\otimes 2}$. See Appendix for a detailed expression of influence function $\Phi_i$.

\paragraph*{}\textbf{Theorem 3:} Under assumptions 1-12, we have
$$\sup_{t\in (0,\tau]}|\hat{F}_1(t|x_e,\bmx_o)-F_1(t|x_e,\bmx_o)|\rightarrow_p 0$$ and $\sqrt{n}(\hat{F}_1(t|x_e,\bmx_o)-F_1(t|x_e,\bmx_o))$
converges weakly to a mean zero Gaussian process with covariance matrix
\bea
K(s,t)&=&(1-F_1(t|x_e,\bmx_o))(1-F_1(s|x_e,\bmx_o))\\
&&[g(t\wedge s)+ts(x_e,\bmx_o)\Omega_{\tau}^{-1}(\Psi_{\tau}+\Sigma_{\tau})\Omega_{\tau}^{-1}(x_e,\bmx_o)^\textsf{T}+(x_e,\bmx_o)\Omega_{\tau}^{-1}(\Gamma(t)s+\Gamma(s)t)].
\eea

\section{Simulation}
We randomly sample $X_I$ from $\{0, 1\}$ with equal probability. $X_o$ and $X_u$ are sampled from independent standard normal distributions. $X_e$ is generated as $X_e=(1,X_I,X_o,X_u)\bmgamma^T$ with $\bmgamma=(0,\gamma_2,0.5,-1)$. By changing $\gamma_2$ from 0.2 to 0.4, we mimic the setting with weak and strong instrumental variables. We vary the sample size $n=100, 400, 1000$. Let $t_0=0.6$ be a fixed time point and denote $t^{\ast}=\min(t, t_0)$. We simulate event time data from the following subdistribution rate
\bea
F(t,\epsilon=1)&=&1-\left\{1-p\left(1-e^{-t}\right)\right\}\exp\{-\beta_e X_e t-\beta_o X_o t-\beta_u X_u t\}\\
F(t,\epsilon=2)&=&\frac{(1-p)\left(1-e^{-t}\right)\exp\{-\beta_e X_e t^{\ast} -\beta_o X_o t^{\ast} - \beta_u X_ut^{\ast}\}}{\exp\{-\beta_e X_e t_0 - \beta_o X_o t_0 - \beta_u X_u t_0\}}
\eea
In this way, we gaurantee that $P(\epsilon=1)=p$ and $P(\epsilon=2)=1-p$ have a sum $F(\infty, \epsilon=1)+F(\infty, \epsilon=2)=1$ and both $F(t,\epsilon=1)$ and $F(t,\epsilon=2)$ are increasing functions. We first generate $\epsilon$ and then generate $T_1$ and $T_2$ given $\epsilon$ by
\bea
F(t|\epsilon=1)&=&\frac{1-\left\{1-p(1-e^{-t})\right\}\exp\{-\beta_e X_e t^{\ast} - \beta_o X_o t^{\ast} -\beta_u X_u t^{\ast}\}}{1-(1-p)\exp\{-\beta_e X_e t_0 - \beta_o X_o t_0 - \beta_u X_u t_0\}}\\
F(t|\epsilon=2)&=&\frac{\exp\{-\beta_e X_e t^{\ast} - \beta_o X_o t^{\ast}- \beta_u X_u t^{\ast}\}}{\exp\{-\beta_e X_e t_0 - \beta_o X_o t_0 -\beta_u X_u t_0\}}
\eea
We set censoring time as exponentially distributed with parameters set to make 30\% and 50\% censoring rate. We set $p=0.8$, $\bmbeta=(0.5,0.2,\beta_3)$. When setting $\beta_3=0$, we have the case of no unmeasured confoundings. When setting $\beta_3=0.2$ and $0.4$, we have the cases with weak and strong unmeasured confoundings. The results are summarized in table 1.

\begin{table}[ht]
\tiny
\caption{Simulation results comparing IV and traditional regression methods in terms of Bias, Standard error (SE), Coverage rate (CR) of 95\% nominal confidence interval under different confounding levels (none/weak/strong) and IV strength (none/weak/strong).}
\centering
\begin{tabular}{|ccc|ccc|ccc|ccc|ccc|}
\hline
\multicolumn{3}{|c|}{} & \multicolumn{6}{|c|}{50\% Censoring} & \multicolumn{6}{|c|}{30\% Censoring}  \\
\hline
\multicolumn{3}{|c|}{} & \multicolumn{3}{|c|}{IV method} & \multicolumn{3}{|c|}{Regression method}  & \multicolumn{3}{|c|}{IV method} & \multicolumn{3}{|c|}{Regression method} \\
\hline
Confounding &IV  & n & Bias & SE & CR & Bias &SE & CR &  Bias & SE & CR & Bias & SE & CR \\
  \hline
None & None & 100 & 5.78 & 37.62 & 96\% & 0.02 & 0.52 & 93\% & 3.96 & 33.22 & 97\% & -0.01 & 0.41 & 94\% \\
None & None & 400 & 5.62 & 47.21 & 97\% & -0.00 & 0.24 & 90\% & 3.17 & 37.01 & 97\% & -0.01 & 0.21 & 92\% \\
None & None & 1000 & -1.75 & 32.13 & 97\% & -0.01 & 0.13 & 95\% & -4.51 & 25.77 & 96\% & -0.01 & 0.12 & 95\% \\
\hline
None & Weak & 100 & -0.08 & 1.64 & 96\% & -0.01 & 0.50 & 92\% & -0.14 & 1.37 & 97\% & -0.03 & 0.41 & 95\% \\
None & Weak & 400 & -0.02 & 0.63 & 98\% & 0.01 & 0.24 & 94\% & 0.02 & 0.54 & 98\% & -0.00 & 0.20 & 92\% \\
None & Weak & 1000 & 0.07 & 0.44 & 93\% & -0.00 & 0.13 & 97\% & 0.06 & 0.39 & 94\% & -0.01 & 0.12 & 94\% \\
\hline
None & Strong & 100 & -0.03 & 0.71 & 94\% & -0.02 & 0.44 & 94\% & -0.06 & 0.57 & 96\% & -0.04 & 0.36 & 95\% \\
None & Strong & 400 & -0.01 & 0.32 & 98\% & 0.00 & 0.20 & 94\% & 0.02 & 0.27 & 98\% & -0.00 & 0.17 & 95\% \\
None & Strong & 1000 & 0.03 & 0.23 & 95\% & 0.00 & 0.12 & 97\% & 0.02 & 0.19 & 96\% & 0.00 & 0.11 & 95\% \\
\hline
Weak & None & 100 & 5.66 & 34.67 & 96\% & -0.22 & 0.55 & 92\% & 3.75 & 32.58 & 97\% & -0.23 & 0.45 & 90\% \\
Weak & None & 400 & 5.61 & 52.28 & 97\% & -0.20 & 0.28 & 84\% & 3.59 & 33.77 & 98\% & -0.21 & 0.23 & 82\% \\
Weak & None & 1000 & -1.56 & 39.24 & 92\% & -0.21 & 0.14 & 73\% & -3.90 & 29.79 & 93\% & -0.21 & 0.13 & 64\% \\
\hline
Weak & Weak & 100 & -0.04 & 1.74 & 95\% & -0.22 & 0.53 & 91\% & -0.08 & 1.41 & 96\% & -0.23 & 0.44 & 89\% \\
Weak & Weak & 400 & -0.04 & 0.69 & 97\% & -0.18 & 0.27 & 83\% & 0.02 & 0.58 & 98\% & -0.18 & 0.23 & 86\% \\
Weak & Weak & 1000 & 0.06 & 0.50 & 93\% & -0.18 & 0.14 & 80\% & 0.05 & 0.43 & 94\% & -0.18 & 0.13 & 69\% \\
\hline
Weak &Strong & 100 & -0.03 & 0.77 & 96\% & -0.18 & 0.46 & 91\% & -0.04 & 0.61 & 98\% & -0.18 & 0.39 & 90\% \\
Weak & Strong & 400 & -0.02 & 0.35 & 97\% & -0.15 & 0.23 & 88\% & 0.01 & 0.29 & 97\% & -0.14 & 0.19 & 88\% \\
Weak & Strong & 1000 & 0.02 & 0.25 & 93\% & -0.13 & 0.13 & 88\% & 0.02 & 0.22 & 95\% & -0.13 & 0.12 & 81\% \\
\hline
Strong & None & 100 & 3.49 & 30.60 & 95\% & -0.45 & 0.58 & 84\% & 2.48 & 19.67 & 96\% & -0.45 & 0.48 & 82\% \\
Strong & None & 400 & 4.01 & 55.30 & 97\% & -0.40 & 0.31 & 67\% & 2.10 & 35.71 & 97\% & -0.41 & 0.26 & 57\% \\
Strong & None & 1000 & -2.21 & 41.40 & 91\% & -0.41 & 0.16 & 28\% & -4.35 & 30.16 & 94\% & -0.41 & 0.15 & 25\% \\
\hline
Strong & Weak & 100 & 0.08 & 2.07 & 96\% & -0.41 & 0.57 & 87\% & 0.02 & 1.71 & 96\% & -0.41 & 0.47 & 84\% \\
Strong & Weak & 400 & -0.02 & 0.74 & 97\% & -0.36 & 0.30 & 68\% & 0.04 & 0.61 & 98\% & -0.36 & 0.24 & 64\% \\
Strong & Weak & 1000 & 0.03 & 0.54 & 91\% & -0.37 & 0.16 & 37\% & 0.05 & 0.47 & 93\% & -0.36 & 0.15 & 27\% \\
\hline
Strong & Strong & 100 & 0.00 & 0.89 & 95\% & -0.32 & 0.52 & 84\% & -0.02 & 0.69 & 96\% & -0.32 & 0.42 & 81\% \\
Strong & Strong & 400 & -0.00 & 0.38 & 96\% & -0.28 & 0.26 & 73\% & 0.02 & 0.31 & 98\% & -0.27 & 0.22 & 70\% \\
Strong & Strong & 1000 & 0.02 & 0.28 & 93\% & -0.27 & 0.15 & 50\% & 0.03 & 0.23 & 94\% & -0.27 & 0.13 & 42\% \\
   \hline
\end{tabular}
\end{table}

From the results, we can see that when IV is strong, the IV method has small bias and correct coverage rate for its 95\% nominal confidence interval. However, when IV becomes weak, the IV estimator performance becomes unsatisfactory, especially in the small sample size case (i.e. $n$=100). When IV assumption does not hold and the instrumental variable is independent of the exposure, the IV method fails (with very large bias and variance) as we expected. When IV is strong, we compare the IV method and the traditional regression method. When there are no unmeasured confoundings, both methods yield small bias and correct coverage rates for their 95\% nominal confidence interval, although the IV estimator is not as efficient as regression based estimator. However, as confounding effects increase, we observe the dramatic increase of bias and smaller coverage rate for regression method while the IV estimator still maintain small bias and the standard error increases very slightly. The coverage rate is approximately at its nominal level.

In practice, we recommend first checking whether the IV is strong or not by looking at some ad-hoc statistics from first stage regression model. For example, the F-statistics$>$10 is often used in econometric literature \cite{r50} as a criteria for strong IV. When an IV is not very strong while the confounding might be small. The balance-variance trade-off might favorate using traditional regression method. So we recommend running both models and compare the results. If the two estimators are very different, then we can see bias in traditional regression method and IV estimator shall be used. If there are no significant difference between the two, then the traditional regression results could be used while the IV results can be considered as sensitivity analysis.

\section{Real transplant data example}

We now apply the proposed instrumental variable analysis method to a real data example. We considered data from the Center for International Blood and Marrow Transplant Research (CIBMTR), on 994 Diffuse Large B Cell Lymphoma (DLBCL) patients, who are 18 to 76 years old, and had an autologous hematopoietic stem cell transplantation (TX) between 1996 and 2003 \cite{r8}. The CIBMTR is comprised of clinical and basic scientists who share data on their blood and bone marrow transplant patients, with the CIBMTR Data Collection Center located at the Medical College of Wisconsin. The CIBMTR has a repository of information regarding the results of transplants at more than 450 transplant centers worldwide. The specific aim of the study was to compare the clinical outcomes between whether rituximab was (+R cohort, N=176) or was not ($-$R cohort, N=818) administered as therapy prior to the transplant. The clinical outcomes compared include relapse/progression of DLBCL, non-relapse mortality (NRM: defined as death without relapse/progression), progression-free survival (PFS) and overall survival (OS). Here relapse/progression and NRM are two competing risks. 355 patients were censored at end of study and censoring distribution is independent of all adjusted covariates by fitting a Cox model for the censoring distribution including all adjusted covariates ($p>0.10$ for each covariate). The CIBMTR study treated the {Year of TX} (2000-2003 versus 1996-1999) as a regular covariate which needs to be adjusted if it is significant in the regression mode, and concluded that pre-TX rituximab did not affect NRM (Hazard Ratio (HR)=0.70, p=0.18) with {Year of TX} to be included in the model (HR=0.63, p=0.05).The risk of relapse/progression was lower in the +R cohort compared to the $-$R cohort (HR=0.67, p=0.004) without {Year of TX} to be included in the final model since it was not significant (HR=0.86, p=0.25).

\begin{table}[ht]
\small
\centering
\caption{Fitting results for CIBMTR study using traditional regression with ``\textsf{Year of TX}" (Model: 1) and without ``\textsf{Year of TX}" (Model: 2) and using ``\textsf{Year of TX}" as an instrumental variable (Model: 3). Adjusted covariates include age, disease status and number of lines of chemotherapy for relapse/progression and age, time from diagnosis to treatment (DX to TX), Karnofsky Score for NRM.}
\begin{tabular}{|c|c|l|c|c|}\hline
Outcome & Model & Variable & Coef. Est. (SE)  & P-value \\\hline
Relapse & 1 & Ritux: Yes vs No & $-.0022$ (0.0016) & 0.1774\\
Relapse & 1 & Age: $\ge 55$ vs $< 55$ & 0.0033 (0.0015) & 0.0320\\
Relapse & 1 & Disease Status: Other vs CR1/PIF & 0.0065 (0.0015)& $<.0001$\\
Relapse & 1 & Prior lines of therapy: $>2$ vs $\le 2$ & 0.0081 (0.0019) & $<.0001$\\
Relapse & 1 & Yr of TX: 2000-03 vs 1996-99 & $-.0035$ (0.0015) & 0.0162\\\hline
Relapse & 2 & Ritux: Yes vs No & $-.0043$ (0.0015) & 0.0036\\
Relapse & 2 & Age: $\ge 55$ vs $< 55$ & 0.0033 (0.0016) & 0.0346\\
Relapse & 2 & Disease Status: Other vs CR1/PIF & 0.0067 (0.0015)& $<.0001$\\
Relapse & 2 & Prior lines of therapy: $>2$ vs $\le 2$ & 0.0083 (0.0019) & $<.0001$\\\hline
Relapse & 3 & Ritux: Yes vs No & $-.0104$ (0.0031) & 0.0007\\
Relapse & 3 & Age: $\ge 55$ vs $< 55$ & 0.0038 (0.0016) & 0.0175\\
Relapse & 3 & Disease Status: Other vs CR1/PIF & 0.0063 (0.0015)& $<.0001$\\
Relapse & 3 & Prior lines of therapy: $>2$ vs $\le 2$ & 0.0088 (0.0019) & $<.0001$\\\hline
NRM & 1 & Ritux: Yes vs No & $-.0006$ (0.0007) & 0.3595 \\
NRM & 1 & Age: $\ge 55$ vs $< 55$ & 0.0015 (0.0006) & 0.0133\\
NRM & 1 & Time from DX to TX: $>1$ vs $\le 1$ Yr & 0.0009 (0.0006) & 0.1163 \\
NRM & 1 & Karnofsky Score: $\ge 90$ vs $<90$ & $-.0013$ (0.0006) & 0.0302 \\
NRM & 1 & Karnofsky Score: Unknown vs $<90$ & 0.0004 (0.0017) & 0.8103 \\
NRM & 1 & Yr of TX: 2000-03 vs 1996-99 & $-.0019$ (0.0006) & 0.0012\\ \hline
NRM & 2 & Ritux: Yes vs No & $-.0017$ (0.0006) & 0.0060 \\
NRM & 2 & Age: $\ge 55$ vs $< 55$ & 0.0014 (0.0006) & 0.0218\\
NRM & 2 & Time from DX to TX: $>1$ vs $\le 1$ Yr & 0.0009 (0.0006) & 0.1307 \\
NRM & 2 & Karnofsky Score: $\ge 90$ vs $<90$ & $-.0013$ (0.0006) & 0.0328 \\
NRM & 2 & Karnofsky Score: Unknown vs $<90$ & $-.0006$ (0.0016) & 0.7263 \\\hline
NRM & 3 & Ritux: Yes vs No & $-.0049$ (0.0012) & $<.0001$ \\
NRM & 3 & Age: $\ge 55$ vs $< 55$ & 0.0018 (0.0006) & 0.0040\\
NRM & 3 & Time from DX to TX: $>1$ vs $\le 1$ Yr & 0.0008 (0.0006) & 0.1390 \\
NRM & 3 & Karnofsky Score: $\ge 90$ vs $<90$ & $-.0014$ (0.0006) & 0.0273 \\
NRM & 3 & Karnofsky Score: Unknown vs $<90$ & 0.0001 (0.0018) & 0.9510 \\\hline
\end{tabular}
\label{tab:2}
\end{table}

\begin{figure}[!h]
\includegraphics[width =5in]{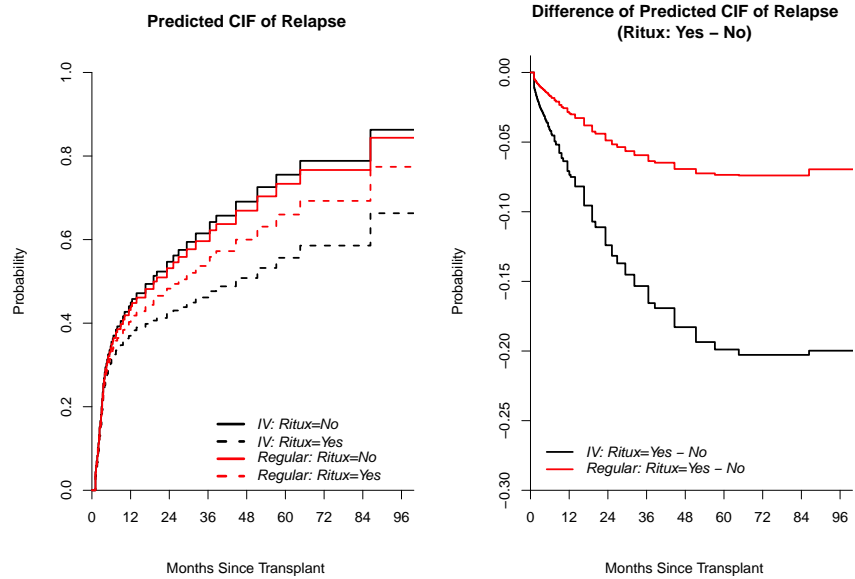}
\caption{Predicted cumulative incidence function of Relapse comparing IV fitting to regular regression analysis}\label{f:rel}
\end{figure}

\begin{figure}[!h]
\includegraphics[width =5in]{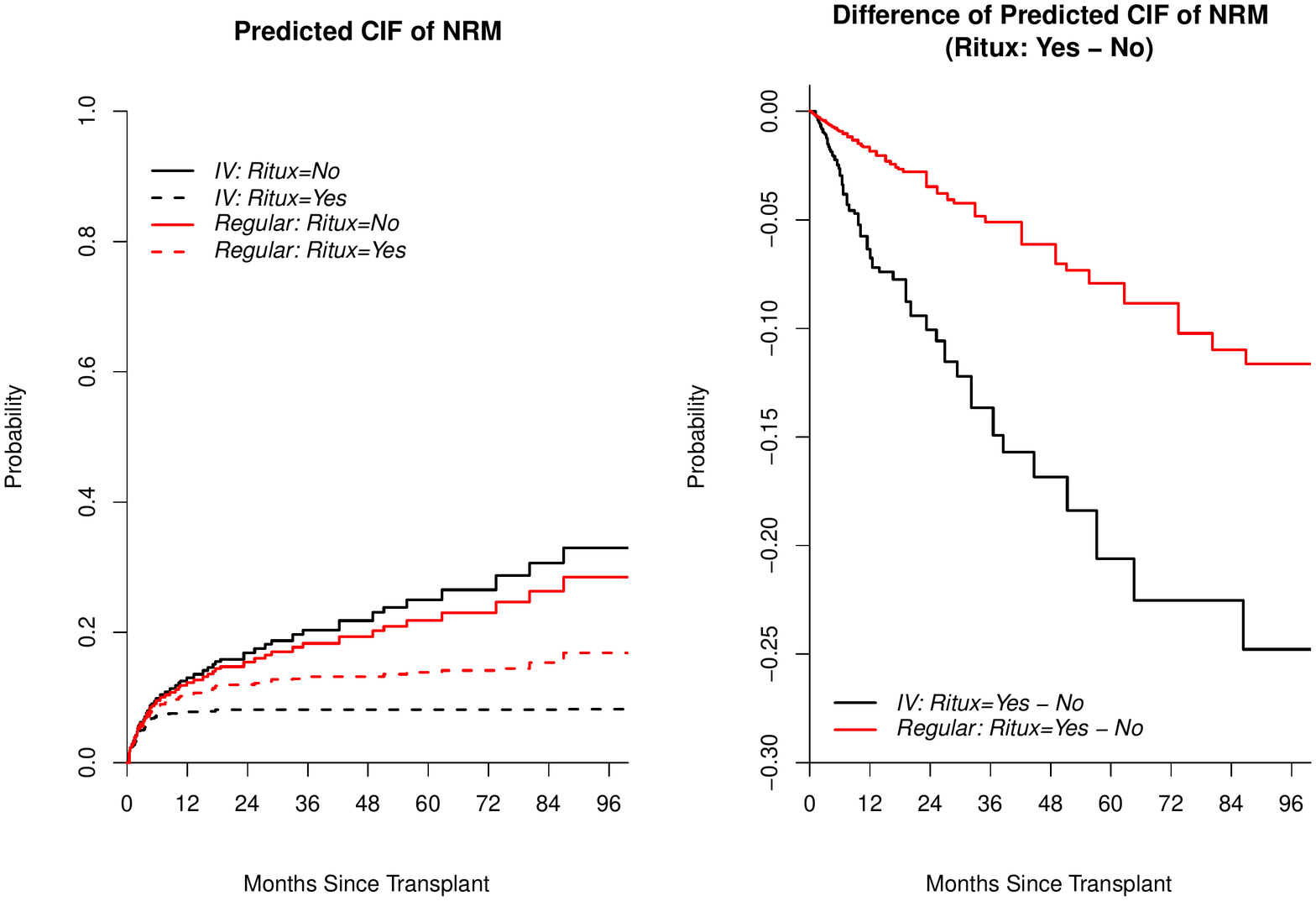}
\caption{Predicted cumulative incidence function of NRM comparing IV fitting to regular regression analysis}\label{f:trm}
\end{figure}

In the CIBMTR study cohort, there were 18\% (174/986) patients who received rituximab in pre-transplant therapy and only 1\% (6/482) had rituximab among those transplanted prior to 1999. Calendar time (\textsf{Year of TX}) was highly associated with the use of rituximab since increase in use rituximab was a result of approval by the US Food and Drug Administrations in 1998. Prior to 1998 Rituximab was available only in clinical trials. Thus, including calendar time (\textsf{Year of TX}) as a regular covariate in a regression analysis could lead to a biased conclusion since we cannot estimate the true effect of rituximab in the presence of \textsf{Year of TX}. Besides observed adjusted covariates (see \cite{r8}), there exist some unobserved covariates, such as molecular subtype of lymphoma (e.g. germinal center vs. activated B cell vs double hit lymphoma) \cite{r3}. These additional prognostic covariates were unknown at the time that these patients were treated and the analysis was done. Such unobserved covariates may affect clinical outcomes and maybe imbalanced between the +R and $-$R cohorts. In the setting of autologous transplantation, transplant related mortality is low ($<3\%$ at 1 year) and there were no major technical improvements during this short time period. Also there were no major clinical practice changes over this short study time period that would have had an impact on transplant in lymphoma except the FDA approval of rituximab. So it is reasonable to assume that the calendar time over such short study time period will not be associated with any of potential unmeasured confounding variables. We use \textsf{Year of TX} (before or after the FDA approval the drug) as an instrumental variable (IV) in proposed IV additive regression model to analyze competing risks data. Risk factors adjusted were \textsf{AGE}, \textsf{Disease Status} and \textsf{Number of Lines of Chemotherapy} for relapse/progression and were \textsf{AGE}, \textsf{Time from diagnosis (DX) to TX} and \textsf{Karnofsky Score} for NRM. The analysis results for fitting a regular regression including \textsf{Year of TX} (Model 1), without \textsf{Year of TX} (Model 2), and IV regression analysis (Model 3) are given in Table \ref{tab:2}. When \textsf{Year of TX} was incorrectly to be included in the regression models, pre-transplant rituximab did not affect both on relapse/progression and NRM since part of the drug effects were counted by the artificial effect of calendar time of \textsf{Year of TX}. Without adjusting \textsf{Year of TX}, estimated rituximab effects were underestimated, where estimated effect size of rituximab by regular regression compared to IV analysis were increased from $-0.0043$ to $-0.0104$ (relatively, $1.42$ times increasing) for relapse/progression and increased from $-0.0017$ to $-0.0049$ (relatively, $1.88$ times increasing) for NRM. Increasing in the size of rituximab effect on the cumulative incidence function (CIF) can be seen in the plot of predicted CIF for +R cohort versus $-$R cohort. At 72 months since TX, the rituximab effect in CIF of relapse/progression were 20\% and 7\% by IV regression analysis and by regular regression analysis, respectively (Figure \ref{f:rel}). Similarly, the rituximab effect in CIF of NRM were 23\% and 8\% by IV regression analysis and by regular regression analysis, respectively (Figure \ref{f:trm}). We observe significant increase in rituximab effect by IV regression modeling. We performed a goodness of fit test to check whether all covariates have constant effect \cite{r32}. The test indicates that the main treatment effect has a constant effect ($p>0.1$), and ``Disease Status'' and ``prior lines of therapy" have time-varying effects for relapse. While for NRM, the main treatment effect has a constant effect and ``Time from DX to TX'', ``Karnofsky Score" and ``missingness of Karnofsky Score" have time-varying effects. We considered fitting more flexible models allowing time-varying effects for covariates that failed the constant effect test. Both models give almost identical results for the main effect ($<1\%$ difference in estimation). Thus, for the illustrate purpose we present all results based on constant effect additive model. Another limitation for the current analysis is that the exposure is binary, while the method require us to fit a linear model of it. To check whether the linear model approximately hold, we check the interaction between the binary instrumental variable and other covariates and did not notice strong interaction, also we performed simulation analysis to check how bias the result could be if the true model is logistic. From table S1, we could see that the non-identity link could lead to some bias in the estimator, however it tends to also increase the variability and the coverage rate is still reasonable.

With calendar time of before or after the FDA approval of rituximab (\textsf{Year of TX}) as an instrumental variable, we confirmed rituximab effect both on relapse/progression and NRM. In conclusion, it is important to identify instrumental variable, and to model and analyze the instrumental variable appropriately in regression analyses.

\section{Discussion}

Only Kaplan-Meier estimate for the censoring distribution is considered through out the paper, which requires a strong assumption that censoring time $C$ is independent of observed covariates. It is clear that this may not be true. He et al. \cite{r30} showed that using covariates adjusted censoring weight can reduce the bias and improve the efficiency in fitting a proportional subdistribution hazards model when the censoring distribution is dependent of some covariates. Thus it is useful to generalize the proposed methods and study the performance by using a regression model for the censoring distribution.

When the exposure is non-gaussian and generalized linear model instead of linear model holds for the exposure model, the mean-variance relationship will cause the variance of $\bmgamma_u^\textsf{T}\bmX_u+e$ depends on $X_I$ and $\bmX_o$ and thus on $\tilde{X}_e$ and introduce potential bias depending on the degree of nonlinearity. When the mean condition holds while the variance condition not, we will have bias in the second stage estimator from the second order approximation. However, if the variance is not very sensitive to the mean or the range of mean is small, the term is still approximately uncorrelated to $\tilde{X}_e$ and thus the bias is small, for example in our real data example.

In this paper, we proposed consistent estimators of efficacy of treatment in the presence of unmeasured confoundings. We showed the asymptotic results of these estimators and studied their performance under finite sample size using simulation. The data example shows that by adjusting unmeasured confounding using instrumental variable, we are able to identify stronger treatment effects. One limitation of this method is that due to the nature of additive models, in certain cases it is hard to guarantee the subdistribution hazard to be positive. Although it is quite clear in our example that the approval from FDA will not directly affect the survival outcome, finding a valid instrumental variable may be challenging in some other settings. The sensitivity analysis is required if we are not sure whether the exclusion restriction assumption holds.

Although we assume time-independent effects in our model, the method can be straightforwardly extended to allow time-varying effects $\bmbeta(t)$ for the two-stage estimator with a more complicated theoretical variance form. In general we can replace the two-stage approach by the following one-stage approach:
$$
0=E[I(T>t,\epsilon=1)\exp\{-\bmbeta^{\textsf{T}}(t)(X_e,\bmX_o)\}(X_I-EX_I|\bmX_o)]
$$
Since $I(T>t,\epsilon=1)$ is not always observable due to the censoring, we can use the IPCW method with weights derived in Zhang \emph{et al.} \cite{r27}. The estimating equation can be written as
$$
0=E[R(t)I(T>t,\epsilon=1)\exp\{-\bmbeta^{\textsf{T}}(t)(X_e,\bmX_o)\}(X_I-EX_I|\bmX_o)]
$$
This allows weaker assumptions (i.e. uncorrelated assumption rather than independence assumption for $X_I$ and outcome conditional on $X_e$, $\bmX_o$) and more efficient results. One the other hand, the advantage of the two-stage approach is computational and it is easy to be implemented using current available softwares.

In conclusion, the two stage least square estimator is a useful method to handle unmeasured confounding variables for competing risk data analysis and it has good finite sample size and asymptotic performance. The two stage approach is easy to implement using available softwares and can be used in various settings including time-varying effect and time-varying covariate $\bmX_o$. The design of measuring useful instrument variable shall provide us better understanding of the treatment effect under potential confounding for compliance.





\appendix


\section*{Appendix}

\subsection*{Proof of Theorem 1:}
We first study the behavior of $S_{1n}(\hat{\gamma},\hat{\Lambda}_c)$. We denote $\tilde{\bmX}_i=(\tilde{X}_{ei},\bmX_{oi})$ and $\hat{\bmX}_i=(\hat{X}_{ei},\bmX_{oi})$. Denote $M^{\ast}_i(t)=N_i^{\ast}(t)-\int_0^t Y_i^{\ast}dH_i(t)$ as an square integrable martingale. We can rewrite $S_{1n}(\hat{\bmgamma},\hat{\Lambda}_c)$ as
\bea
S_{1n}(\hat{\bmgamma},\hat{\Lambda}_c)&=&n^{-1}\sum_{i=1}^n\int_0^\tau \left[\hat{\bmX}_{i}^\textsf{T}-\frac{\sum_{j=1}^n\hat{\bmX}_j^\textsf{T}w_j(t)\hat{Y}^2_j(t)}{\sum_{j=1}^nw_j(t)\hat{Y}^2_j(t)}\right]w_i(t)\hat{Y}_i(t)d\hat{N}_i(t)\\
&=&n^{-1}\sum_{i=1}^n\int_0^\tau \left[\hat{\bmX}_i^\textsf{T}-\frac{\sum_{j=1}^n\hat{\bmX}_j^\textsf{T}w_j(t)\hat{Y}^2_j(t)}{\sum_{j=1}^nw_j(t)\hat{Y}^2_j(t)}\right]w_i(t)\hat{Y}_i(t)dM^{\ast}_i(t)\\
&&+n^{-1}\sum_{i=1}^n\int_0^\tau \left[\hat{\bmX}_i^\textsf{T}-\frac{\sum_{j=1}^n\hat{\bmX}_j^\textsf{T}w_j(t)\hat{Y}^2_j(t)}{\sum_{j=1}^nw_j(t)\hat{Y}^2_j(t)}\right]w_i(t)\hat{Y}_i(t)(\hat{R}_i(t)-R_i(t))dN_i(t)\\
&=&S_{11n}+S_{12n}
\eea
For Kaplan Meier estimator, we have
\bea
\hat{R}_i(t)-R_i(t)=-\frac{G(t)I(T_i<t)}{G(T_i)}\sum_{j=1}^n\int_{T_i}^t \frac{dM^c_{j}(u)}{\sum_{k=1}^n I(T_k\geq u)}
\eea
where $M^c_j(u)=I(T_i\leq u, \epsilon_i=0)-\int_0^u I(T_i\geq t)d\Lambda^c(t)$. Due to the consistency of $\hat{\gamma}$ and uniform convergence of $\hat{G}(\cdot)$, we have
\bea
S_{12n}=n^{-1}\sum_{i=1}^n\int_0^\tau \left[\tilde{\bmX}_i^\textsf{T}-\frac{\sum_{j=1}^n\tilde{\bmX}_j^\textsf{T}w_j(t)Y^{\ast 2}_j(t)}{\sum_{j=1}^nw_j(t)Y^{\ast 2}_j(t)}\right]w_i(t)Y^{\ast}_i(t)(\hat{R}_i(t)-R_i(t))dN_i(t)+o_p(1)
\eea
So we have $S_{12n}$ is an average of martingale integrals with respect to the censoring filtration $\mathcal{F}^c(u)=\{I(T_i\leq u, \epsilon_i=0), I(T_i\geq t),\bmX_i,i=1,\cdots,n,t\leq u\}$. Under regularity condition, this term is dominated by a bounded function and converge in probability to 0 uniformly in a compact neighborhood of $\bmbeta_0$ \cite{r16}. Due to the consistency of $\hat{\bmgamma}$ and uniform convergence of $\hat{G}(\cdot)$, we also have
\bea
&&\left[\hat{\bmX}_i^\textsf{T}-\frac{\sum_{j=1}^n\hat{\bmX}_j^\textsf{T}w_j(t)\hat{Y}^2_j(t)}{\sum_{j=1}^nw_j(t)\hat{Y}^2_j(t)}\right]w_i(t)\hat{Y}_i(t)=\left[\tilde{\bmX}_i^\textsf{T}-\frac{\sum_{j=1}^n\tilde{\bmX}_j^\textsf{T}w_j(t)Y^{\ast 2}_j(t)}{\sum_{j=1}^nw_j(t)Y^{\ast 2}_j(t)}\right]w_i(t)Y^{\ast}_i(t)+o_p(1)
\eea
So we have
\bea
S_{11n}=n^{-1}\sum_{i=1}^n\int_0^\tau \left[\tilde{\bmX}_i^\textsf{T}-\frac{\sum_{j=1}^n\tilde{\bmX}_j^\textsf{T}w_j(t)Y^{\ast 2}_j(t)}{\sum_{j=1}^nw_j(t)Y^{\ast 2}_j(t)}\right]w_i(t)Y^{\ast}_i(t)dM^{\ast}_i(t)+o_p(1)
\eea

Due to the consistency of $\hat{\gamma}$ and uniform convergence of $\hat{G}(\cdot)$, we can approximate $S_{2n}(\hat{\gamma},\hat{\Lambda}_c)$ by
\bea
S_{2n}(\hat{\bmgamma},\hat{\Lambda}_c)=n^{-1}\sum_{i=1}^n\int_{0}^{\tau}\left[\tilde{\bmX}_i^\textsf{T}-\frac{\sum_{j=1}^n\tilde{\bmX}_j^\textsf{T}w_j(t)Y^{\ast 2}_j(t)}{\sum_{j=1}^nw_j(t)Y^{\ast 2}_j(t)}\right]^{\otimes 2}w_i(t)Y^{\ast 2}_i(t)dt+o_p(1)
\eea

Based on the consistency results for traditional additive competing risk model without instrumental variable, $\tilde{\bmbeta}=S_{2n}(\bmgamma,\Lambda_c)^{-1}S_{1n}(\bmgamma,\Lambda_c)$ converge to $\beta_0$. Since we have shown that our estimator $\hat{\bmbeta}-\tilde{\bmbeta}=o_p(1)$, so we obtain the consistency for $\hat{\beta}$.

\subsection*{Proof of Theorem 2:}
Denote $U_n(\bmbeta)=S_{2n}\bmbeta-S_{1n}$, we have $S_{2n}$ converge to positive definite matrix $\Omega_{\tau}$. To obtain distribution of $\hat{\bmbeta}$, we know
\bea
\sqrt{n}(\hat{\bmbeta}-\bmbeta_0)=A_{\tau}^{-1}\sqrt{n}U_n(\bmbeta_0)+o_p(1)
\eea
So, we just need to provide an expansion of $\sqrt{n}U_n(\bmbeta_0)$. We have
\bea
&&\sqrt{n}U_n(\bmbeta_0)\\
&=&n^{-1/2}\sum_{i=1}^n\int_{0}^{\tau}\left[\bmX_{IOE}(\bmX_{Ioi}-\frac{\sum_{j=1}^n\bmX_{Ioj}w_j(t)\hat{Y}^2_j(t)}{\sum_{j=1}^nw_j(t)\hat{Y}^2_j(t)})\right]w_i(t)\hat{Y}_i(t)[d\hat{N}_i(t)-\hat{Y}_i(t)\bmbeta_0^\textsf{T}\bmX_{IOE}\bmX_{Ioi} dt]\\
&=&n^{-1/2}\sum_{i=1}^n\int_{0}^{\tau}\left[\bmX_{IOE}(\bmX_{Ioi}-\frac{\sum_{j=1}^n\bmX_{Ioj}w_j(t)\hat{Y}^2_j(t)}{\sum_{j=1}^nw_j(t)\hat{Y}^2_j(t)})\right]w_i(t)\hat{Y}_i(t)dM_i^{\ast}(t)\\
&&+n^{-1/2}\sum_{i=1}^n\int_{0}^{\tau}-\left[\bmX_{IOE}(\bmX_{Ioi}-\frac{\sum_{j=1}^n\bmX_{Ioj}w_j(t)\hat{Y}^2_j(t)}{\sum_{j=1}^nw_j(t)\hat{Y}^2_j(t)})\right]w_i(t)\hat{Y}_i^2(t)\bmbeta_0^\textsf{T}\bmX_{IOE}\bmX_{Ioi} dt]\\
&&+n^{-1/2}\sum_{i=1}^n\int_{0}^{\tau}\left[\bmX_{IOE}(\bmX_{Ioi}-\frac{\sum_{j=1}^n\bmX_{Ioj}w_j(t)\hat{Y}^2_j(t)}{\sum_{j=1}^nw_j(t)\hat{Y}^2_j(t)})\right]w_i(t)\hat{Y}_i(t)[\hat{R}_i(t)-R_i(t)]dM_i(t)\\
&=&n^{-1/2}\sum_{i=1}^n\int_{0}^{\tau}\left[\bmX_{IOE}(\bmX_{Ioi}-\frac{\sum_{j=1}^n\bmX_{Ioj}w_j(t)Y^{\ast 2}_j(t)}{\sum_{j=1}^nw_j(t)Y^{\ast 2}_j(t)})\right]w_i(t)Y^{\ast}_i(t)dM_i^{\ast}(t)\\
&&+n^{-1/2}\sum_{i=1}^n\int_{0}^{\tau}-\left[\bmX_{IOE}(\bmX_{Ioi}-\frac{\sum_{j=1}^n\bmX_{Ioj}w_j(t)Y^{\ast 2}_j(t)}{\sum_{j=1}^nw_j(t)Y^{\ast 2}_j(t)})\right]w_i(t)Y_i^{\ast 2}(t)\bmbeta_0^\textsf{T}\bmX_{IOE}\bmX_{Ioi} dt]\\
&&+n^{-1/2}\sum_{i=1}^n\int_{0}^{\tau}\left[\bmX_{IOE}(\bmX_{Ioi}-\frac{\sum_{j=1}^n\bmX_{Ioj}w_j(t)Y^{\ast 2}_j(t)}{\sum_{j=1}^nw_j(t)Y^{\ast 2}_j(t)})\right]w_i(t)Y^{\ast}_i(t)[\hat{R}_i(t)-R_i(t)]dM_i(t)+o_p(1)\\
&&=S_{21n}+S_{22n}+S_{23n}
\eea
For the first term, it is in the form of martingale integration and we have
\bea
S_{21n}=n^{-1/2}\sum_{i=1}^n\Phi_{1i}(\tau)+o_p(1)
\eea
and
\bea
Var(S_{21n})=n^{-1}\sum_{i=1}^n\int_{0}^{\tau}\left[\bmX_{IOE}(\bmX_{Ioi}-\frac{\sum_{j=1}^nX_{Ioj}w_j(t)Y^{\ast 2}_j(t)}{\sum_{j=1}^nw_j(t)Y^{\ast 2}_j(t)})\right]^{\otimes 2}w_i(t)Y^{\ast}_i(t)dN^{\ast}_i(t)+o_p(1)
\eea
converge to $\Psi_{\tau}$ with probability 1.

For the second term, we have
\bea
S_{22n}&=&n^{-1/2}\sum_{i=1}^n\int_{0}^{\tau}-\left[\bmX_{IOE}(\bmX_{Ioi}-\frac{\sum_{j=1}^n\bmX_{Ioj}w_j(t)Y^{\ast 2}_j(t)}{\sum_{j=1}^nw_j(t)Y^{\ast 2}_j(t)})\right]w_i(t)Y_i^{\ast 2}(t)\beta_0^\textsf{T}\bmX_{IOE}\bmX_{Ioi} dt]\\
&=&n^{-1/2}\sum_{i=1}^n\int_{0}^{\tau}-\left[\bmX_{IOE}(\bmX_{Ioi}-\frac{\sum_{j=1}^n\bmX_{Ioj}w_j(t)Y^{\ast 2}_j(t)}{\sum_{j=1}^nw_j(t)Y^{\ast 2}_j(t)})\right]w_i(t)Y_i^{\ast 2}(t)\bmX_{Ioi}^\textsf{T}\bmX_{IOE}^\textsf{T}\bmbeta_0 dt]\\
&=&n^{-1/2}\sum_{i=1}^n\Phi_{2i}(\tau)+o_p(1) \\
\eea
Notice $\bmX_{IOE}^\textsf{T}\bmbeta_0=\beta_{0e}[\bmX_{Io}^\textsf{T}\bmX_{Io}]^{-1}\bmX_{Io}^\textsf{T}X_e$, we can approximate $S_{22n}$ by
\bea
n^{-1/2}\sum_{i=1}^n\int_{0}^{\tau}-\beta_{0e}\left[\bmX_{IOE}(\bmX_{Ioi}-\frac{\sum_{j=1}^n\bmX_{Ioj}w_j(t)Y^{\ast 2}_j(t)}{\sum_{j=1}^nw_j(t)Y^{\ast 2}_j(t)})\right]\bmX_{Ioi}^\textsf{T}[\bmX_{Io}^\textsf{T}\bmX_{Io}]^{-1}\bmX_{Io}^\textsf{T}X_ew_i(t)Y_i^{\ast 2}(t)dt]\\
\eea
\bea
Var(S_{22n})&=&\frac{\sigma^2\beta_{0e}^2}{n}\sum_{i=1}^n\int_{0}^{\tau}\left[\bmX_{IOE}(\bmX_{Ioi}-\frac{\sum_{j=1}^n\bmX_{Ioj}w_j(t)Y^{\ast 2}_j(t)}{\sum_{j=1}^nw_j(t)Y^{\ast 2}_j(t)})\right]\bmX_{Ioi}^\textsf{T}[\bmX_{Io}^\textsf{T}\bmX_{Io}]^{-1}\bmX_{Ioi}\\
&&\times \left[X_{IOE}(\bmX_{Ioi}-\frac{\sum_{j=1}^n\bmX_{Ioj}w_j(t)Y^{\ast 2}_j(t)}{\sum_{j=1}^nw_j(t)Y^{\ast 2}_j(t)})\right]^\textsf{T}w_i(t)Y_i^{\ast 2}(t)dt+o_p(1)
\eea
converge to $\Sigma_{\tau}$ with probability 1. The covariance between $S_{11n}$ and $S_{22n}$ converge to 0 with probability 1 as in \cite{r13}.

Now we look at the last term $S_{23n}$. Under same argument as \cite{r9}, the empirical process and Lenglart's inequality \cite{r12} with there approximations provide us
\bea
\sqrt{n}S_{23n}=n^{-1/2}\sum_{i=1}^n\int_0^\tau \frac{q(u)}{\pi(u)}dM_i^c(u)+o_p(1)=n^{-1/2}\sum_{i=1}^n\Phi_{3i}(\tau),
\eea
where
\bea
q(u)&=&-lim_{n\rightarrow \infty} n^{-1}\sum_{i=1}^n\int_0^\tau \left[\bmX_{IOE}(\bmX_{Ioi}-\frac{\sum_{j=1}^n\tilde{\bmX}_i^\textsf{T}w_i(t)Y^{\ast 2}_i(t)}{\sum_{j=1}^nw_i(t)Y^{\ast 2}_i(t)})\right]w_i(t)Y^{\ast}_i(t)dM_i(t) I(t\geq u\geq T_i)\\
\pi(u)&=&lim_{n\rightarrow \infty} n^{-1}\sum_{i=1}^n I(T_i\geq u)
\eea
So the third term variance is 0 and thus is ignorable comparing to the first two terms. As weighted estimating equation approach, the third term is asymptotically uncorrelated with first two terms and thus we obtain the asymptotic variance of the $\hat{beta}$ as $\Omega_{\tau}^{-1}(\Psi_{\tau}+\Sigma_{\tau})\Omega_{\tau}^{-1}$ with influence function $\Phi_i=\Phi_{1i}+\Phi_{2i}+\Phi_{3i}$.

\subsection*{Proof of Theorem 3:}
We will first study the performance of $-\log (1-\hat{F}_1(t|X_e=x_e,\bmX_o=\bmx_o))=\hat{H}_0^{\ast}(t)+t\hat{\beta}^\textsf{T}[x_e,\bmx_o]$. We can write this as
\bea
&&n^{1/2}(\hat{H}_0^{\ast}(t)+t\hat{\bmbeta}^\textsf{T}[x_e,\bmx_o]-H_0^{\ast}(t)-t\bmbeta_0^\textsf{T}[x_e,\bmx_o])\\
&=&n^{1/2}(\int_0^{t}\frac{\sum_{i=1}^nw_i(s)\hat{Y}_i(s)[d\hat{N}_i(s)-\hat{Y}_i(s)\hat{\bmbeta}^\textsf{T}\bmX_{IOE}\bmX_{Ioi}ds]}{\sum_{i=1}^nw_i(s)\hat{Y}^2_i(s)}-H_0^{\ast}(t)+t(\hat{\bmbeta}-\bmbeta_0)^\textsf{T}[x_e,\bmx_o])\\
&=&n^{1/2}(\int_0^{t}\frac{\sum_{i=1}^nw_i(s)Y^{\ast}_i(s)[dN^{\ast}_i(s)-Y^{\ast}_i(s)\hat{\bmbeta}^\textsf{T}\bmX_{IOE}\bmX_{Ioi}ds]}{\sum_{i=1}^nw_i(s)Y^{\ast 2}_i(s)}-H_0^{\ast}(t)+t(\hat{\bmbeta}-\bmbeta_0)^\textsf{T}[x_e,\bmx_o])+o_p(1)\\
\eea
The first term can be written as $n^{-1/2}\sum_{i=1}^n\zeta_{1i}$ with covariance between two time $t$ and $t'$ as
\bea
\int_0^{t\wedge t'}\frac{n^{-1}\sum_{i=1}^nw_i^2(s)Y^{2 \ast}_i(s)dN^{\ast}_i(s)}{[n^{-1}\sum_{i=1}^nw_i(s)Y^{\ast 2}_i(s)]^2}
\eea
converge to $g(t\wedge t')$ in probability uniformly.

The second term has the covariance between two time $t$ and $t'$ as
\bea
tt'(x_e,\bmx_o)\Omega_{\tau}^{-1}(\Psi_{\tau}+\Sigma_{\tau})\Omega_{\tau}^{-1}(x_e,\bmx_o)^\textsf{T}
\eea

The covariance between two time $t$ and $t'$ for first term and second term converge in probability to
\bea
(x_e,\bmx_o)\Omega_{\tau}^{-1}\Gamma(t)t'
\eea

So the Martingale Central limit theory provide the asymptotic of  $-\log (1-\hat{F}_1(t|X_e=x_e,\bmX_o=\bmx_o))$. Using $\Delta-$method, we can obtian the results in theorem 3.

\section*{Supplement}
We studied the effect of first stage model misspecification using parameter estimated from the real data example (i.e., the survival parameter from model 3 and the logistic regression parameter from regression $X_e$ on $X_I$ and $X_O$) with $n=986$. The unmeasured confounder is simulated from $U(-0.5,0.5)$ and the confounding strength are set the same as the weakest covariate's effect. Then we use the parameter to simulate $X_e$, $Y$, $\delta$ and $\delta\epsilon$. The result shows that although small bias exists, due to the increase in variability, the coverage rate is still reasonable.
\begin{table}[ht]
\small
\caption*{\textbf{Table S1}: Simulation results studying IV performance under model misspecification.}
\centering
\begin{tabular}{|cccc|cccc|}
\hline
\multicolumn{4}{|c|}{Relapse} & \multicolumn{4}{|c|}{NRM}   \\
\hline
 True Effect & Bias & SE & CR &  True Effec & Bias &SE & CR \\
  \hline
0.0104 & 0.0031 & 0.0039 & 94\% &0.0049 & 0.0001 & 0.0012 & 98\%\\
   \hline
\end{tabular}
\end{table}
\end{document}